\documentclass[aps,pre,twocolumn,10pt,showpacs,superscriptaddress,nofootinbib]{revtex4}
\usepackage{amsmath,amssymb}
\usepackage{graphicx}
\usepackage[usenames]{color}
\usepackage{rotating,array,tabularx,booktabs,colortbl}
\usepackage{relsize}
\newcolumntype{Y}{>{\centering\arraybackslash}X}
\definecolor{violet}{rgb}{0.5,0,0.5}

\newcommand{\dif}{\mathrm{d}}%
\newcommand{\fdif}{\operatorname{\delta}}
\newcommand{\Fdif}[2]{\frac{\fdif\!#1}{\fdif\!#2}}
\newcommand{\Laplace}{\boldsymbol{\triangle}}%
\newcommand{\ie}{i.\,e.}%
\newcommand{\eg}{e.\,g.}%
\newcommand{\ZT}[1]{\textquotedblleft#1\textquotedblright}%
\newcommand{\iDelta}{\mathlarger{\blacktriangle}}%

\graphicspath{{img/}}

\begin{document}
\title{Structure and dynamics of interfaces between two coexisting liquid crystalline phases}

\author{Simon Praetorius}\author{Axel Voigt}  
\affiliation{Institute of Scientific Computing, Technical University Dresden, D-01062 Dresden, Germany}

\author{Raphael Wittkowski}
\affiliation{School of Physics and Astronomy, University of Edinburgh, Edinburgh, EH9 3JZ, United Kingdom}

\author{Hartmut L{\"o}wen}
\affiliation{Institut f{\"u}r Theoretische Physik II, Weiche Materie,
Heinrich-Heine-Universit{\"a}t D{\"u}sseldorf, D-40225 D{\"u}sseldorf, Germany}

\date{\today}

\begin{abstract}
The phase-field-crystal model is used to access the structure and thermodynamics of 
interfaces between two coexisting liquid crystalline phases in two spatial dimensions. 
Depending on the model parameters there is a variety of possible coexistences  
between two liquid crystalline phases including a plastic triangular crystal (PTC). 
Here, we calculate numerically the profiles for the mean density and the nematic order tensor across 
the interface for isotropic-PTC and columnar-PTC respectively smectic A-PTC coexistence.
As a general finding, the width of the interface with respect to the nematic order parameter characterizing 
the orientational order is larger than the width of the mean density interface. 
In approaching the interface from the PTC side, at first the mean density goes down and then the 
nematic order parameter follows. 
The relative shift of the two profiles can be larger than a full lattice constant of the plastic crystal.
Finally, we also present numerical results for the dynamic relaxation of an initial 
order-parameter profile towards its equilibrium interfacial profile. 
Our predictions for the interfacial profiles can in principle be verified in real-space experiments of colloidal dispersions. 

\end{abstract}

 
\maketitle

\section{Introduction}
Liquid crystals typically composed of anisometric molecules or colloidal particles form interesting 
mesophases, which are neither completely liquid nor crystalline \cite{deGennesP1995}. 
The simultaneous presence of translational and rotational degrees of freedom gives 
rise to phases, which exhibit a different degree of ordering for the translational and orientational 
order \cite{Frenkel1991,VroegeL1992}. 
Rotator solids or plastic crystals, for instance, are translationally ordered but orientationally disordered, 
while nematics, on the other hand, possess orientational order in the absence of translational order.
Clearly, there is also the fully disordered isotropic phase and the fully ordered crystalline phase,  
but there are even more intermediate liquid crystalline phases (such as, for example, smectic-A and columnar phases) 
with different degrees of translational and orientational order that are stable for appropriate thermodynamic conditions. 

It is a formidable task of statistical physics to predict the existence and stability of the different 
liquid crystalline phases for a given interaction as a function of mean density and temperature. This has, for example, 
been done by computer simulations of simple model systems \cite{BolhuisF1997,BrownAdRM1998} and also by molecular 
density functional theory \cite{PoniewierskiH1988,GrafL1999} and more phenomenological approaches \cite{GrafL1999b}.
Typically, phase diagrams of liquid crystals exhibit regions, where two phases of different kind of ordering coexist. 
At equal pressure, chemical potential, and temperature, coexistence implies that there is a stable interface between 
the two coexisting phases. 
In mean-field theories (which neglect interfacial capillary wave undulations), the interface has a characteristic width 
of typically several particle sizes and exhibits profiles of the mean density and of the degree of orientational order 
depending on the spatial coordinate perpendicular to the interface. 
For the liquid-solid interface, see, for example, Refs.\ 
\cite{Curtin1987,Curtin1989,LoewenBW1989,OhnesorgeLW1991,MarrG1993,OhnesorgeLW1994}.

While there has been a large effort to explore the gas-liquid \cite{Evans1979} and liquid-solid \cite{Woodruff1980} interfaces 
of spherical particles 
(see also Refs.\ \cite{BinderM2000,HoytAK2001,DavidchackML2006,ZykovaTimanRHB2009,ZykovaTimanHB2010,RozasH2011,HaertelOREHL2012}), 
much less effort has been devoted to the particle-resolved structure and thermodynamics of the interface between two coexisting 
\textit{liquid crystalline} phases.
Extensive studies have been performed for the isotropic-nematic interface, which has been accessed by experiment, computer simulation, 
and theory (see, for example, 
Refs.\ \cite{McDonaldAS2001,VelascoMS2002,BierHD2004,VinkS2005,vanderBeekEtAl2006,WolfsheimerTSvRS2006,ReichDvRS2007,UllrichASV2010,VerhoeffOvdSL2011}),  
but there are fewer considerations of the isotropic-smectic \cite{MederosS1992,SomozaMS1995,Blanc2001,DogicF2001} and the nematic-smectic interface 
\cite{OsipovSR2003}. 
However, to the best of our knowledge there is no investigation of an interface, where one of the coexisting phases is 
plastic or fully crystalline.
This is, of course, a nontrivial task, since there is a complex dependence of the interface structure on the 
(relative) orientation of the two phases. Even for the isotropic-crystal coexistence there is a complex
orientational dependence culminating in Wulff's construction for the equilibrium crystal shape \cite{HaertelOREHL2012}.
Nevertheless, it is important to have information about the interface, since nucleation and growth phenomena of a metastable phase in a stable phase 
occur via interfaces \cite{DogicF2001,SchillingF2004b,VerhoeffL2012}.

In this paper, we close this gap and study liquid crystalline interfaces also for crystalline phases.
We use a phase-field-crystal (PFC) model, which is a minimal model to describe freezing for isotropic 
particles on the molecular (\ie, interparticle) scale \cite{ElderKHG2002,JaatinenAN2010,EmmerichEtAl2012} 
and can be justified from microscopic density functional theory \cite{ElderPBSG2007,vanTeeffelenBVL2009,EmmerichEtAl2012}. 
The traditional PFC model \cite{ElderKHG2002} was later generalized to anisotropic particles in two \cite{Loewen2010} and 
three \cite{WittkowskiLB2010} spatial dimensions allowing for liquid crystalline phases. 
The generalized theory is formulated in terms of three order-parameter fields, 
namely the reduced translational density $\psi(\vec{r})$, 
the local nematic order parameter $S(\vec{r})$, and the mean orientational direction $\hat{n}(\vec{r})$ 
that is also called the \ZT{nematic director}. 
While the traditional PFC model \cite{ElderKHG2002} has two free parameters, the liquid crystalline PFC model in 
two dimensions \cite{Loewen2010} has five independent couplings. 
This widely opens the parameter space for the occurrence of several liquid crystalline phases including 
nematic, columnar, smectic-A, plastic crystalline, and orientationally ordered crystalline phases. 
Recent numerical studies \cite{AchimWL2011} of the liquid crystalline PFC model in two spatial dimensions have shown that 
a variety of phase coexistences occur as a function of the model parameters. 
Therefore, the liquid crystalline PFC model \cite{Loewen2010} provides a simple and direct avenue to access the interface structure, 
which still incorporates the correct physics.

As a result, we find that the width of the interface with respect to the nematic order parameter is larger than 
the width of the mean density interface. 
In approaching the interface from the plastic crystalline side, at first the mean density goes down and then the nematic order parameter follows. 
The relative shift of these two profiles can be larger than a full lattice constant of the plastic crystal.
Finally, we also present numerical results for the dynamic relaxation of an initial order-parameter profile towards 
its equilibrium interfacial profile. Our results can in principle be verified in 
real-space experiments of colloidal dispersions, which can be confined to monolayers 
\cite{RoordavDPGvBK2004,TurkovicDF2005,WargackiPV2008,CrassousDPMDHDS2012}. A transient non-monotonic behavior
of the conserved mean-density profiles occurs, which is much more pronounced than non-monotonicities 
in the non-conserved orientational order profile.

The paper is organized as follows: after the presentation of a suitable PFC model for liquid crystals in Sec.\ \ref{sec:PFC}, 
we describe a numerical method for the solution of the dynamical PFC equations in Sec.\ \ref{sec:Numerik}.
Our results obtained by numerical calculations are discussed in Sec.\ \ref{sec:Ergebnisse}.
Finally, we conclude in Sec.\ \ref{sec:Zusammenfassung}.

\section{\label{sec:PFC}PFC model for liquid crystals in two spatial dimensions}
A PFC model for apolar\footnote{We neglect a possible macroscopic polarization.} 
liquid crystals in two spatial dimensions was given in 
Refs.\ \cite{Loewen2010,AchimWL2011,WittkowskiLB2011,WittkowskiLB2011b,EmmerichEtAl2012}. 
It describes the static properties and dynamical behavior of a liquid crystalline system in terms of two 
dimensionless order-parameter fields: 
the reduced translational density $\psi(\vec{r},t)$ and the symmetric and traceless nematic tensor $Q_{ij}(\vec{r},t)$ 
with position $\vec{r}=(x,y)$ and time $t$. 
For liquid crystalline particles with a symmetry axis, the nematic tensor can be parametrized as  
\begin{equation}
Q_{ij}(\vec{r},t)=S(\vec{r},t)\Big(n_{i}(\vec{r},t)n_{j}(\vec{r},t)-\frac{1}{2}\delta_{ij}\Big) 
\end{equation}
with the nematic order parameter $S(\vec{r},t)$ and the (normalized) nematic director $\hat{n}(\vec{r},t)=(n_{1},n_{2})$ 
(see Refs.\ \cite{Loewen2010,AchimWL2011,WittkowskiLB2011}).

\subsection{Static free-energy functional}
The static properties of a liquid crystalline system are described by a free-energy functional $\mathcal{F}[\psi,Q_{ij}]$, 
which is minimized with respect to $\psi(\vec{r})$ and $Q_{ij}(\vec{r})$ in thermodynamic equilibrium. 
After an appropriate rescaling of the length and energy scales, this free-energy functional obtains the 
dimensionless form\footnote{Einstein's sum convention is used throughout this paper. Notice that powers of indexed quantities  
involve repeated indices and thus summation, \ie, for example, $Q^{2}_{ij}\equiv Q_{ij}Q_{ij}\equiv\sum_{i,j}Q_{ij}Q_{ij}$.} \cite{WittkowskiLB2011b}  
\begin{equation}%
\begin{split}%
\mathcal{F}[\psi,Q_{ij}]&=\!\int\!\!\dif^{2}r\,\bigg(\!-\frac{\psi^{3}}{3}+\frac{\psi^{4}}{6}
+(\psi-1)\frac{\psi Q^{2}_{kl}}{4}\\
&\hspace{-3mm}+\frac{Q^{2}_{kl}Q^{2}_{mn}}{64}+A_{1}\psi^{2}
+A_{2}\psi(\Laplace+\Laplace^{2})\psi \\[0.5mm]
&\hspace{-3mm}+B_{3}(\partial_{k}\psi)(\partial_{l}Q_{kl})+D_{1}Q^{2}_{kl}+D_{2}(\partial_{l}Q_{kl})^{2}\bigg)
\end{split}%
\label{eq:PFCs}%
\end{equation}%
with the Laplace operator $\Laplace\equiv\partial^{2}_{k}$ and the five dimensionless coupling parameters 
$A_{1}$, $A_{2}$, $B_{3}$, $D_{1}$, and $D_{2}$.

\subsection{Dynamical equations}
The corresponding dynamical equations of $\psi(\vec{r},t)$ and $Q_{ij}(\vec{r},t)$ can be derived from 
classical dynamical density functional theory \cite{WittkowskiL2011} 
and are given by \cite{WittkowskiLB2011b} 
{\allowdisplaybreaks
\begin{align}%
\begin{split}%
\dot{\psi} + \partial_{i} J^{\psi}_{i} &= 0 \;, 
\end{split}\label{eq:PFCdI}\\
\begin{split}
\dot{Q}_{ij} + \Phi^{Q}_{ij}           &= 0 
\end{split}\label{eq:PFCdII}%
\end{align}}%
with the dimensionless current $J^{\psi}_{i}(\vec{r},t)$ and the dimensionless quasi-current $\Phi^{Q}_{ij}(\vec{r},t)$. 
In constant-mobility approximation, this current and quasi-current are given by \cite{EmmerichEtAl2012}
{\allowdisplaybreaks
\begin{align}%
\begin{split}%
\!\!\!\!\!\!J^{\psi}_{i}  = &- 2\alpha_{1}(\partial_{i}\psi^{\natural})
-2\alpha_{3}(\partial_{j}Q^{\natural}_{ij}) \;, 
\end{split}\label{eq:Jpsi}\\[3pt]
\begin{split}%
\!\!\!\!\!\!\Phi^{Q}_{ij} = &-4\alpha_{1}(\Laplace Q^{\natural}_{ij})
-2\alpha_{3}\big(2(\partial_{i}\partial_{j}\psi^{\natural})
-\delta_{ij}(\Laplace\psi^{\natural})\big) \!\!\!\!\!\! \\
&+8\alpha_{4}Q^{\natural}_{ij} 
\end{split}\label{eq:PhiQ}%
\end{align}}%
with the three dimensionless mobility parameters $\alpha_{1}$, $\alpha_{3}$, and $\alpha_{4}$ and the thermodynamic conjugates 
\begin{equation}
\psi^{\natural}=\Fdif{\mathcal{F}}{\psi}\;,\qquad
Q_{ij}^{\natural}=\Fdif{\mathcal{F}}{Q_{ij}} 
\end{equation}
of $\psi(\vec{r},t)$ and $Q_{ij}(\vec{r},t)$, respectively. 
The thermodynamic conjugates follow directly from the free-energy functional \eqref{eq:PFCs} by functional differentiation:
{\allowdisplaybreaks%
\begin{align}%
\begin{split}%
\psi^{\natural} = &- \psi^{2} + \frac{2}{3}\psi^{3} + (2\psi-1)\frac{Q^{2}_{ij}}{4} +2A_{1}\psi \\
&+ 2A_{2}(\Laplace+\Laplace^{2})\psi -B_{3}(\partial_{i}\partial_{j}Q_{ij}) \;,
\end{split}\label{eq:psiR}\\%
%%%
\begin{split}%
Q^{\natural}_{ij} = &\,\psi(\psi-1)Q_{ij} + \frac{Q_{ij}Q^{2}_{kl}}{8} \\
&-B_{3}\big(2(\partial_{i}\partial_{j}\psi)-\delta_{ij}\Laplace\psi\big) + 4D_{1}Q_{ij} \\[2pt]
&-2D_{2}\:\!\partial_{k}\big(\partial_{i}Q_{kj}+\partial_{j}Q_{ki}-\delta_{ij}(\partial_{l}Q_{kl})\big) \;. 
\end{split}\label{eq:QR}%
\end{align}}%
For a comparison of the dimensionless rescaled parameters in Eqs.\ \eqref{eq:PFCs}, \eqref{eq:Jpsi}, and \eqref{eq:PhiQ} with   
the corresponding parameters in the notation of Refs.\ \cite{AchimWL2011,WittkowskiLB2011b,EmmerichEtAl2012}, 
see appendix \ref{A:Parameter}.

\section{\label{sec:Numerik}Numerical solution of the PFC model}
By inserting Eqs.\ \eqref{eq:psiR} and \eqref{eq:QR} into Eqs.\ \eqref{eq:Jpsi} and \eqref{eq:PhiQ} one obtains for 
the dynamics \eqref{eq:PFCdI} and \eqref{eq:PFCdII} a system of six coupled nonlinear partial differential equations. 
In order to solve this system numerically, we decouple and linearize it. 
A simplification is possible due to the symmetry and tracelessness of the nematic tensor.
Defining the variables $q_i\equiv Q_{i,1}$ and $q_i^\natural\equiv Q_{i,1}^\natural$, 
we can write the system of dynamical equations as
\begin{equation}
\begin{split}
\dot{\psi} &= 2\alpha_1\Laplace\psi^\natural + 2\alpha_3 \iDelta_i q^\natural_i \;, \\
\dot{q}_i &= 4\alpha_1\Laplace q_i^\natural - 8\alpha_4 q_i^\natural+2\alpha_3\iDelta_i\psi^\natural
\end{split}
\label{eq:dyn_eqn}%
\end{equation}
with the operator 
$\iDelta=(\partial_1\partial_1 - \partial_2\partial_2, 2\partial_1\partial_2)$ 
that is related to the Cauchy-Riemann operator.
The thermodynamic conjugates reformulated in the new variables read 
\begin{equation}
\begin{split}
\psi^\natural &= \omega_\psi(\psi,\vec{q})+2A_1\psi+2A_2(\Laplace+\Laplace^2)\psi-B_3\iDelta_i q_i \;, \\
q^\natural_i &= \omega_q(\psi,\vec{q})_i+4D_1 q_i-2D_2\Laplace q_i-B_3\iDelta_i\psi
\end{split}
\label{eq:thermodyn_conj}\raisetag{4.5ex}%
\end{equation}
with the polynomials 
\begin{equation}
\begin{split}
\omega_\psi(\psi,\vec{q}) &= -\psi^2+\frac{2}{3}\psi^3+\frac{1}{2}(2\psi-1)q_i^2 \;, \\
\omega_q(\psi,\vec{q})_i &= \psi(\psi-1)q_i + \frac{1}{4}q_i q_j^2 \;.
\end{split}
\label{eq:polynomials_w}%
\end{equation}
We could discretize in time using a semi-implicit Euler discretization. 
Let $t_1,t_2,t_3\dotsc$ be a sequence of time steps. Defining 
$\psi_n\equiv \psi(\vec{r},t_n)$, $q_{i,n}\equiv q_i(\vec{r},t_n)$, and $\tau_{n}=t_{n+1}-t_{n}$,
we obtain (by treating some terms explicitly and others implicitly) the decoupled systems
\begin{gather}
\frac{\psi_{n+1}}{\tau_n} - 2\alpha_1\Laplace\psi^\natural_{n+1} = \frac{\psi_{n}}{\tau_n} 
+ 2\alpha_3\iDelta_i q^\natural_{i,n} \;, \\
\begin{split}
\psi^\natural_{n+1} &- 2\big(A_1-A_2(\Laplace+\Laplace^2)\big)\psi_{n+1}  \\ 
&= \omega_\psi(\psi_{n+1},\vec{q}_n) -B_3\iDelta_i q_{i,n} 
\end{split}
\label{eq:euler_discretization}%
\end{gather}
and
\begin{equation}
\begin{split}
\frac{q_{i,n+1}}{\tau_n}- 4(\alpha_1\Laplace-2\alpha_4) q^\natural_{i,n+1} 
&=\frac{q_{i,n}}{\tau_n} + 2\alpha_3\iDelta_i\psi^\natural_{n} \;, \\
q^\natural_{i,n+1}-2(2D_1-D_2\Laplace) q_{i,n+1}  &= \omega_q(\psi_n,\vec{q}_{n+1}) \\
&\quad\:\! -B_3\iDelta_i\psi_n \;.
\end{split}
\label{eq:euler_discretization2}%
\end{equation}
Linearizing $\omega_\psi(\psi,\vec{q})$ and $\omega_q(\psi,\vec{q})_i$ around the old time step $t_{n}$, two linear systems 
can be solved one after the other for all $n$. 
The linearizations of the polynomials \eqref{eq:polynomials_w} read
\begin{equation}
\begin{split}
\omega_\psi(\psi_n, \psi_{n+1}, \vec{q}_n) &= \psi_{n+1}\big(\!-2\psi_n+2\psi_n^2+q_{i,n}^2\big) \\
&\quad\:\! + \psi_n^2-\frac{4}{3}\psi_n^3-\frac{q_{i,n}^2}{2} \;, \\
\omega_q(\psi_n, \vec{q}_n, \vec{q}_{n+1})_i &= \frac{1}{4}q_{i,n+1}\big(4\psi_n(\psi_n-1)+q_{j,n}^2\big) \\
&\quad\:\! +\frac{1}{2}q_{i,n}q_{j,n}q_{j,n+1} -\frac{1}{2}q_{i,n}q_{j,n}^2 \;. 
\end{split}
\label{eq:w_linearized}%
\end{equation}
However, such a simple time stepping scheme with constant step size turns out to be impractical, if stationary configurations 
have to be obtained in the simulations. We therefore used a higher-order embedded Rosenbrock scheme 
with an adequate step size control for the time discretization. 
A detailed description of this scheme concerning some numerical issues will be given elsewhere.
For the discretization in space we used the finite element method.

\section{\label{sec:Ergebnisse}Results}
\begin{figure}[ht]
\centering
\includegraphics[width=\linewidth]{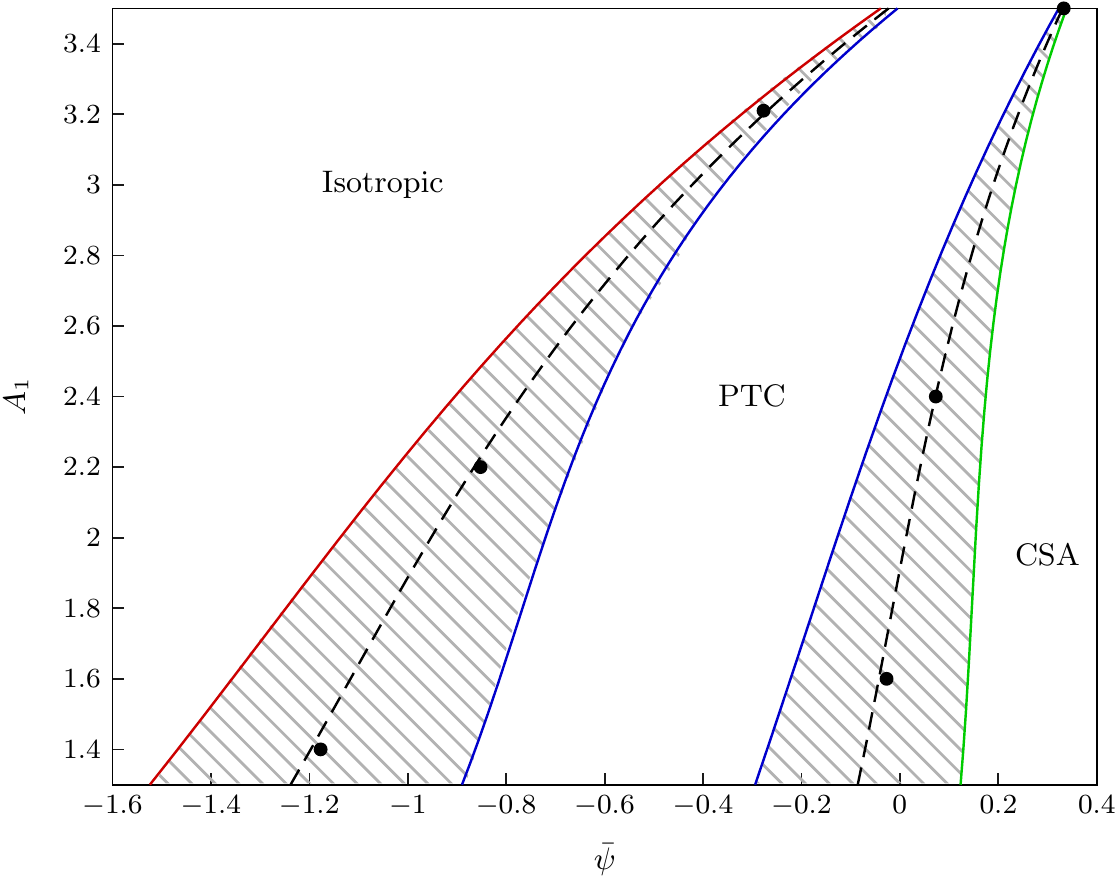}%
\caption{\label{fig:pd}Phase diagram with coexistence regions for the mean density $\bar{\psi}\in[-1.6,0.4]$ and the 
parameters $A_{1}\in[1.3,3.5]$, $A_{2}=14$, $B_{3}=-0.4$, $D_{1}=1$, and $D_{2}=0.8$. 
Three different liquid crystalline phases are realized: isotropic, columnar/smectic A (CSA), 
and plastic triangular crystalline (PTC). 
The coexistence regions (shaded areas) are calculated using Maxwell's double tangent construction. 
The black dashed lines in the coexistence regions indicate the intersection lines of the energy curves of the 
two adjacent phases. Six black circles indicate certain parameter combinations for which detailed calculations were 
performed (see Figs.\ \ref{fig:width}, \ref{fig:distance}, and \ref{fig:interfaceI}-\ref{fig:interfaceIII}).}
\end{figure}

We first restrict ourselves to certain parameter combinations, which allow for several liquid crystalline coexistences.
In detail, we fix the parameters $A_{2}=14$, $B_{3}=-0.4$, $D_{1}=1$, and $D_{2}=0.8$, but vary the parameter $A_{1}$ 
(which corresponds to some formal temperature in the context of mean-field theories) and the 
reduced mean density $\bar{\psi}$.\footnote{The parameters in the dynamical equations \eqref{eq:PFCdI}-\eqref{eq:PhiQ} are always chosen 
to be $\alpha_{1}=\alpha_{3}=\alpha_{4}=1$. Clearly, the stationary results do not depend on their particular values.} 
The resulting equilibrium bulk phase diagram is shown in Fig.\ \ref{fig:pd} in consistency with earlier data \cite{AchimWL2011}. 
In the parameter range of $A_{1}$ and $\bar{\psi}$ shown, the phase diagram exhibits three stable 
liquid crystalline phases, namely, the isotropic phase, a plastic triangular crystal (PTC)\footnote{The 
plastic triangular crystal in phase diagram \ref{fig:pd} is called 
\ZT{plastic triangular crystal 2} (PTC2) in Ref.\ \cite{AchimWL2011}.}, and a columnar phase.
As we consider two spatial dimensions here, a columnar phase is indistinguishable from a smectic A phase,
therefore we call the latter columnar/smectic A (CSA)\footnote{This columnar/smectic A (CSA) phase is called 
\ZT{C/SA phase} in Ref.\ \cite{AchimWL2011}.} phase. 
The coexistence regions, as obtained by a Maxwell double tangent construction,
are depicted by the shaded area in Fig.\ \ref{fig:pd}. 
We selected in total six different coexistence conditions as labeled by black circles in Fig.\ \ref{fig:pd}, 
which correspond to three isotropic-PTC and three CSA-PTC coexistence situations serving as basic reference situations 
for our subsequent investigations.

\begin{figure}[ht]
\centering
\includegraphics[width=\linewidth]{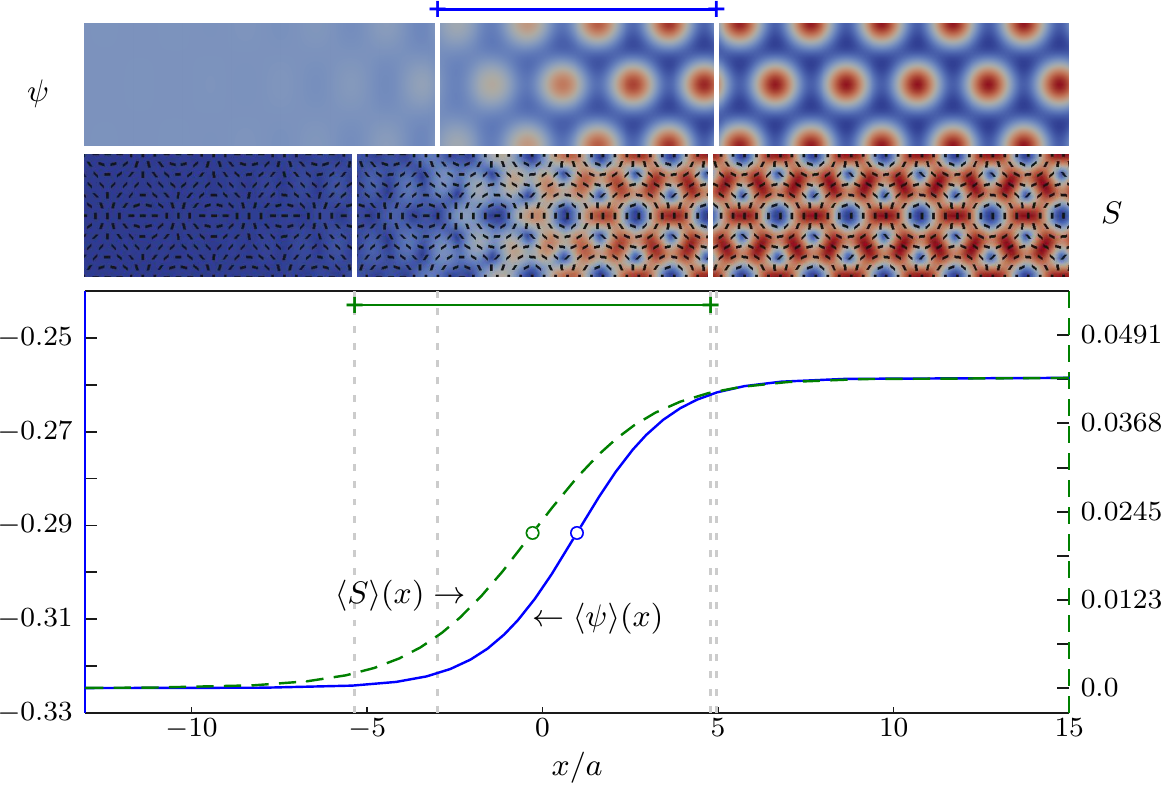}%
\caption{\label{fig:interface}Top: two contour plots for $\psi(x,y)$ and $S(x,y)$ at 
an isotropic-PTC coexistence with $A_1=3.21$ and $\bar{\psi}=-0.3$ 
(the other parameters are the same as in Fig.\ \ref{fig:pd}). $\hat{n}(\vec{r})$ is represented by short black lines 
that are superimposed to the lower contour plot.
Bottom: averaged density $\langle\psi\rangle(x)$ (left ordinate) and averaged nematic order parameter 
$\langle S\rangle(x)$ (right ordinate).   
The $x$-direction is chosen perpendicular to the interface, while the $y$-axis is parallel to the interface. 
The averaged quantities are defined by $\langle f\rangle(x)=\int\!\dif y'\int^{x+a}_{x-a}\!\dif x'\, f(x',y')$ 
for $f\in\{\psi,S\}$ with the width of the stripes $2a=4\pi/(k\sqrt{3})$ and $k=1/\sqrt{2}$.}
\end{figure}

A typical example for an isotropic-PTC interfacial profile is presented in Fig.\ \ref{fig:interface} for the (10)-orientation 
of the hexagonal crystal. 
In the bulk PTC phase, there are periodic peaks in the full density profile $\psi(x,y)$ 
at the crystal lattice positions, shown as a contour plot in Fig.\ \ref{fig:interface}.
The typical standard deviation of these peaks (the so-called Lindemann parameter) is 
pretty large with about $27\%$ of the lattice constant. The corresponding orientational ordering
as embodied in the nematic tensor is complicated and exhibits topological defects in the Wigner-Seitz cell
of the lattice, see Refs.\ \cite{AchimWL2011,CremerML2012} for a more detailed discussion. 
The mean orientational unit vector field $\hat{n}(x,y)$ as obtained by the direction of the eigenvector
of the nematic tensor corresponding to the highest eigenvalue, is sketched by short black lines in Fig.\ \ref{fig:interface}.
The largest eigenvalue itself, the scalar nematic order parameter field $S(x,y)$, is also presented as a contour plot 
in Fig.\ \ref{fig:interface}.
In the isotropic phase, on the other hand, the density field is constant and the nematic order parameter vanishes.
In between there is an interfacial region with laterally averaged profiles $\langle\psi\rangle(x)$ and 
$\langle S\rangle(x)$ with $x$ denoting the direction perpendicular to the interface 
(see caption of Fig.\ \ref{fig:interface}).

We define a typical interface width of an order parameter profile $f(x,y)\in\{\psi(x,y),S(x,y)\}$ 
as the distance of the positions, where a \texttt{tanh}-approximation of $\langle f\rangle(x)$ attains the values 
$0.95\langle f\rangle(-\infty)+0.05\langle f\rangle(\infty)$ and 
$0.05\langle f\rangle(-\infty)+0.95\langle f\rangle(\infty)$, respectively. 
These widths for $\psi(x,y)$ and $S(x,y)$ are indicated in Fig.\ \ref{fig:interface}. 
Remarkably, the width of the density profile is significantly smaller than the width of the orientational profile.
The position, where the \texttt{tanh}-approximation of an averaged field 
$\langle f\rangle(x)$ with $f\in\{\psi,S\}$ attains the value $(\langle f\rangle(-\infty)+\langle f\rangle(\infty))/2$ 
can be taken as a natural location $\xi(f)$ of the interface with respect to this field.
Interestingly, as revealed in Fig.\ \ref{fig:interface}, the location of the averaged density profile $\langle\psi\rangle(x)$ 
and the averaged orientational profile $\langle S\rangle(x)$ do not coincide.
The location of the orientational profile is more shifted towards the isotropic phase than the location of the density profile.
This means that coming from the isotropic side, at first the nematic order builds up and then the density follows. 
This finding is reminiscent to the fluid-crystal interface of systems of spherical particles \cite{Lutsko2010,HaertelOREHL2012}, 
which can be described by a two-order-parameter description involving the conserved mean density and 
a non-conserved crystallinity \cite{LoewenBW1989,LoewenBW1990}. 
Coming from the fluid side, also in this case, the non-conserved crystallinity starts to grow first and the density follows.

\begin{figure}[ht]
\centering
\includegraphics[width=\linewidth]{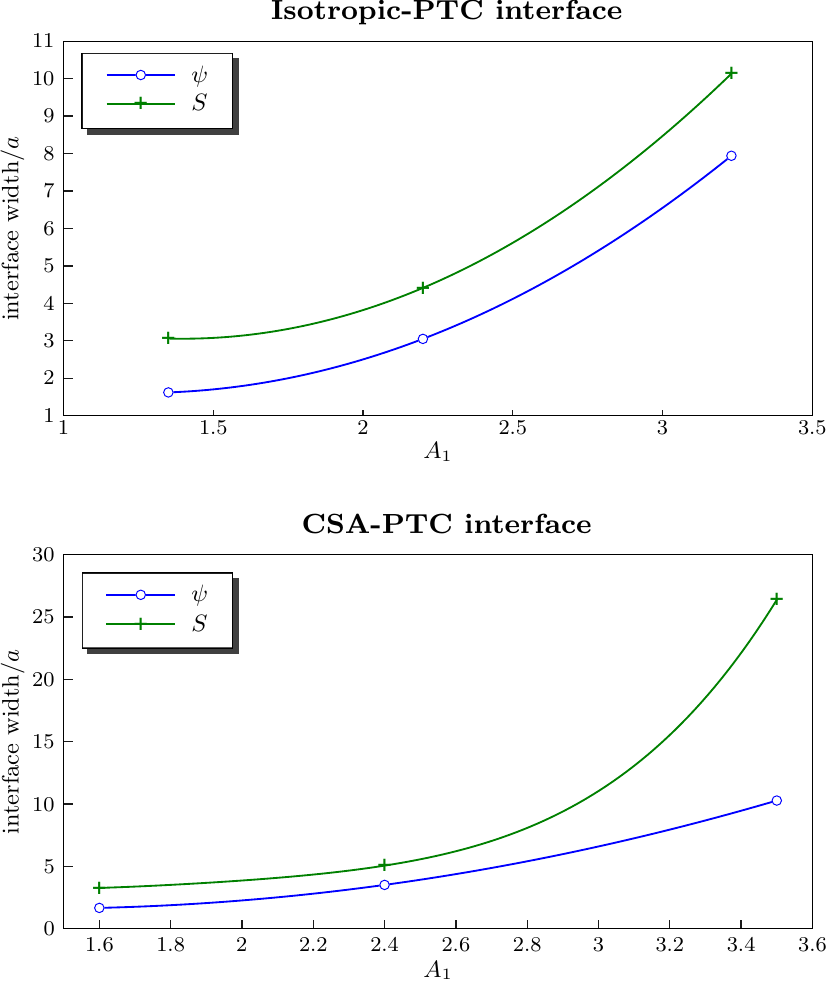}%
\caption{\label{fig:width}Widths of the interfaces of $\psi(\vec{r})$ and $S(\vec{r})$ in dependence of $A_{1}$. 
The parameters are $(\bar{\psi},A_{1})\in\{(-0.05,1.6),(0.05,2.4),(0.31,3.5)\}$ and for the rest as in 
Fig.\ \ref{fig:pd}. Notice that the presented data correspond to the six points highlighted by black circles in Fig.\ \ref{fig:pd} 
and that they are connected by spline interpolation. 
In the lower plot the stripes of the CSA phase are oriented perpendicular to the interface (see Fig.\ \ref{fig:interfaceII}).}
\end{figure}

\begin{figure}[ht]
\centering
\includegraphics[width=\linewidth]{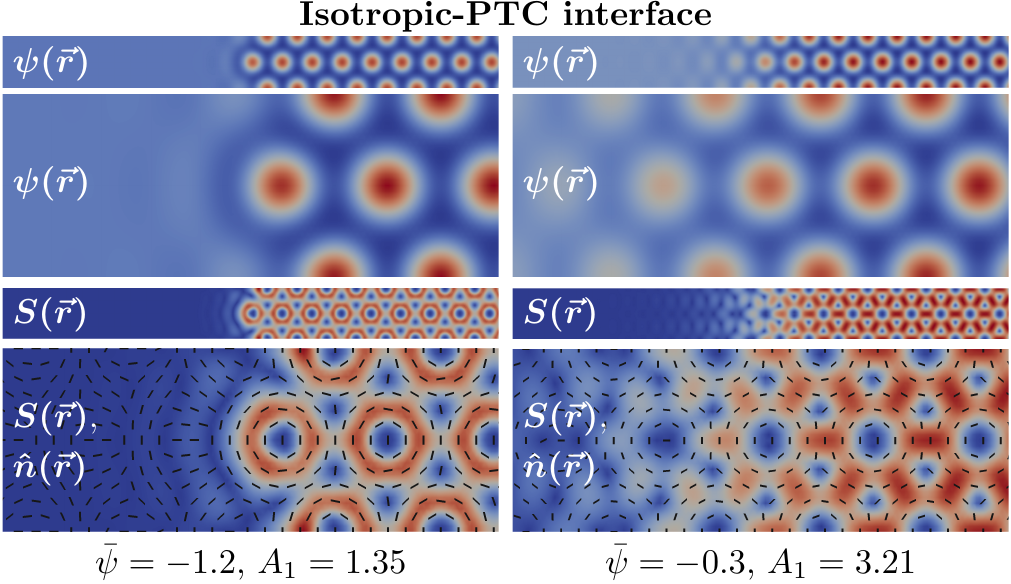}%
\caption{\label{fig:interfaceI}Interface of the isotropic-PTC phase coexistence for the same parameters as in 
Fig.\ \ref{fig:pd}. The plots show the translational density $\psi(\vec{r})$ and the nematic order parameter $S(\vec{r})$ both for a large area 
and for a close-up view of the interface, where blue and red indicate low and high values, respectively. 
In addition, the director field $\hat{n}(\vec{r})$ is represented by short black lines that are superimposed to the lowest plots.} 
\end{figure}

We have further studied the dependence of the interface widths on the parameter $A_{1}$. As $A_{1}$ is increased, the
coexistence comes closer to a critical point where the interfacial widths diverge. 
This trend is documented in Fig.\ \ref{fig:width}. 
The upper plot in Fig.\ \ref{fig:width} also shows that the width of the orientational order-parameter profile is larger than that of the 
density interface over the full range of $A_{1}$. 
All trends are the same for different parameter combinations for the isotropic-PTC interface, as documented by Fig.\ \ref{fig:interfaceI}.

\begin{figure}[ht]
\centering
\includegraphics[width=\linewidth]{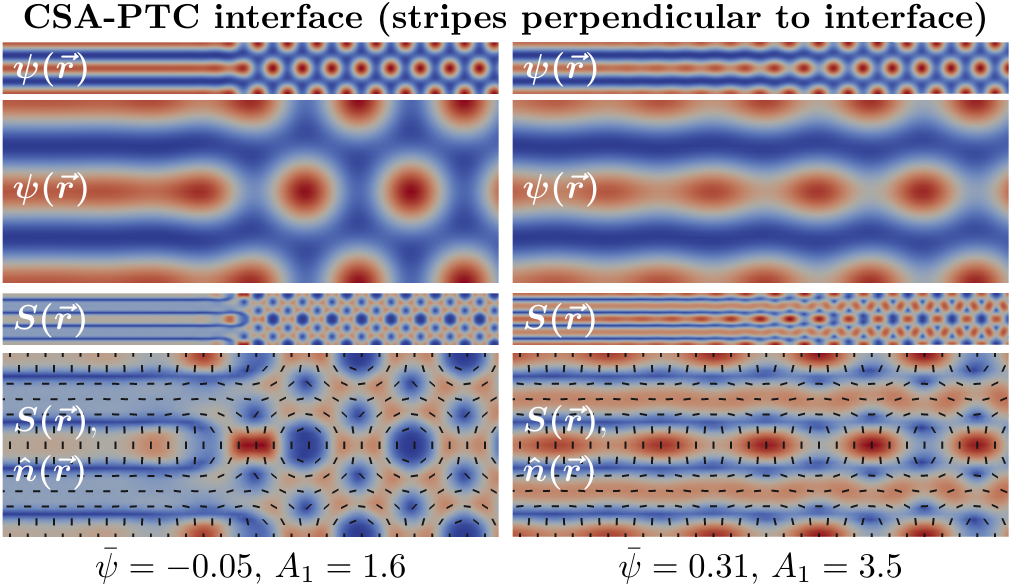}%
\caption{\label{fig:interfaceII}The same as in Fig.\ \ref{fig:interfaceI}, but now for the CSA-PTC interface.
Note that the stripes of the CSA phase are oriented perpendicular to the interface.}
\end{figure}

\begin{figure}[ht]
\centering
\includegraphics[width=\linewidth]{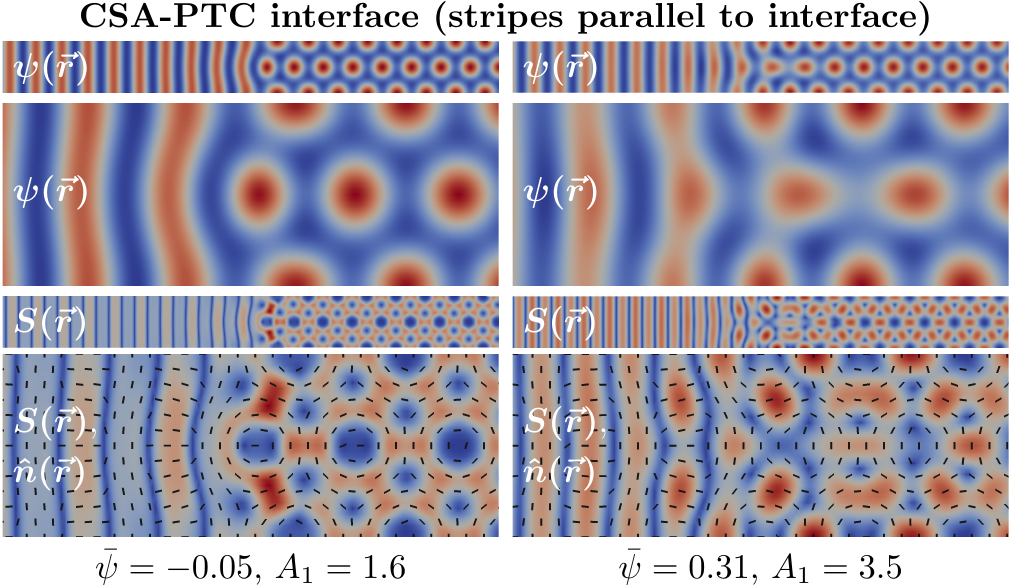}%
\caption{\label{fig:interfaceIII}The same as in Fig.\ \ref{fig:interfaceII}, but now for a CSA-PTC interface, 
where the stripes of the CSA phase are oriented parallel to the interface.}
\end{figure}

Next, we consider the coexistence between the PTC and the CSA phase. In this case, the interface structure 
depends on the relative orientations of the two phases. While we fix the orientation of the PTC phase in the (10)-direction, 
we consider here two possibilities of the column direction relative to the interface, namely perpendicular and parallel.
For these two different relative orientations, the order-parameter fields are given in 
Figs.\ \ref{fig:interfaceII} and \ref{fig:interfaceIII} for two different parameter combinations of coexistence.
For perpendicular column direction (see Fig.\ \ref{fig:interfaceII}), the density field reveals that the columns end 
at a lattice density peak. This implies that the degeneracy of the column positions is broken by the presence of the 
crystal, which pins the transversal columnar order by the interface.
Along the columns away from the interface, there are still some density undulations in $x$-direction.
For parallel column direction (see Fig.\ \ref{fig:interfaceIII}), on the other hand, there is a nontrivial density field 
across the interface insofar as the columns are significantly bent in the presence of the crystalline peaks, \ie, 
the crystal induces a systematic undulation of the neighboring columns. The amplitude of this undulation decreases farer 
away from the interface position. Likewise, along the columns there is a periodic density modulation in $y$-direction 
induced by the crystalline peaks nearby.

\begin{figure}[ht]
\centering
\includegraphics[width=\linewidth]{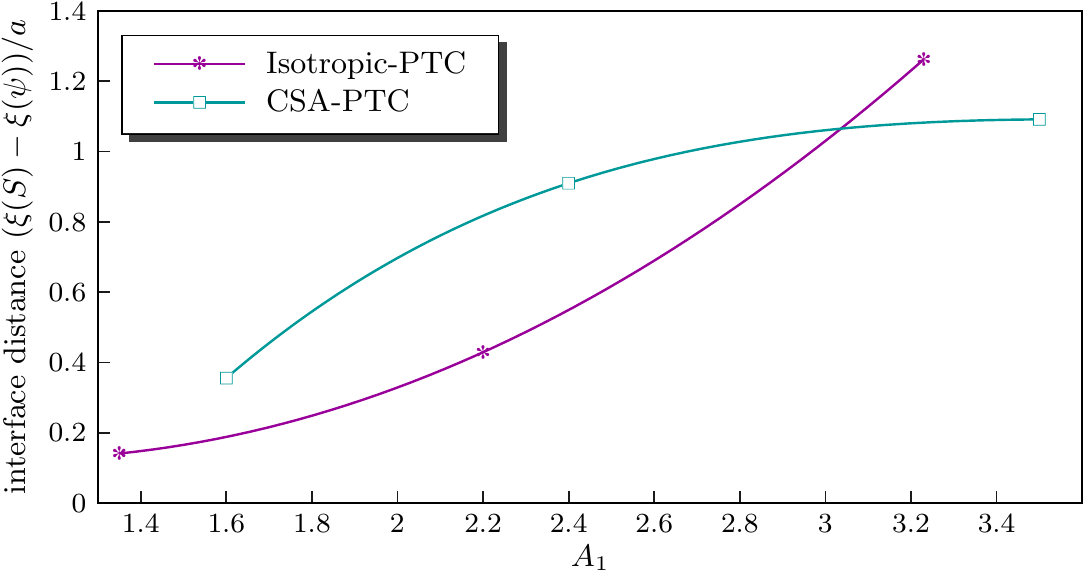}%
\caption{\label{fig:distance}Distance $\Delta \xi(\psi,S) = \xi(S)-\xi(\psi)$ of the interfaces of 
$\psi(\vec{r})$ and $S(\vec{r})$ in dependence of $A_{1}$ (we always consider the transition from the crystalline to the 
non-crystalline phase). The parameters are the same as in Fig.\ \ref{fig:width}  
and the stripes of the CSA phase are again oriented perpendicular to the interface (see Fig.\ \ref{fig:interfaceII}).}
\end{figure}

Results for the interfacial width, similarly defined as in the previous case, are shown in the lower plot in Fig.\ \ref{fig:width}, 
where the same trends are observed as for the isotropic-PTC interface [see the upper plot in Fig.\ \ref{fig:width}]. 
The width of the orientational interface is considerably larger than that for the density profile and there is a strong 
dependence on the parameter $A_{1}$ with huge interfacial widths, where the parameters are close to criticality. 
Like the isotropic-PTC interface, the interface position of the density profile is more in the PTC-phase than the 
interface position of the orientational profile, which is more in the coexisting CSA phase (see Fig.\ \ref{fig:distance}). 
The shift in the two interface positions depends on the parameters, as shown in Fig.\ \ref{fig:distance}.

\begin{figure}[ht]
\centering
\includegraphics[width=\linewidth]{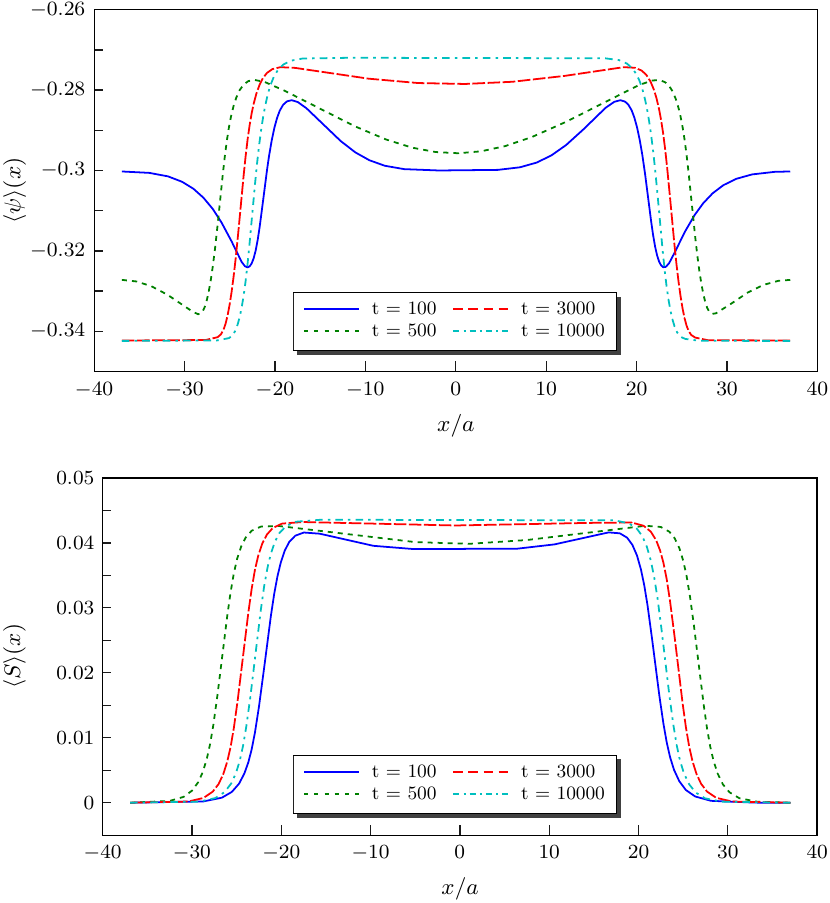}%
\caption{\label{fig:dynamics}Time evolution of the averaged order parameters $\langle\psi\rangle(x)$ (top) 
and $\langle S\rangle(x)$ (bottom) for an isotropic-PTC coexistence. The parameters are $\alpha_{1}=\alpha_{3}=\alpha_{4}=1$,  
$\bar{\psi}=-0.3$, $A_1=3.21$ and for the rest as in Fig.\ \ref{fig:pd}. Snapshots are taken at times $t = 100,500,3000$, and $10000$. 
At time $t=0$ the averaged translational density is constant ($\langle\psi\rangle(x)=-0.3$) and the 
averaged nematic order parameter $\langle S\rangle(x)$ is a smeared Heaviside step function.}
\end{figure}

Finally, we show some results on the dynamical evolution of the interfacial profiles based on the physical dynamics 
described in Eqs.\ \eqref{eq:PFCdI} and \eqref{eq:PFCdII}. 
It is important to note that the density is a conserved order parameter, while the nematic ordering is non-conserved. 
We plot an example of the interface relaxation towards equilibrium for a prescribed starting profile in 
Fig.\ \ref{fig:dynamics}. 
The orientational order-parameter field is a smeared Heaviside step function, while the density is constant. 
Similar set-ups for interfacial kinetics have been studied earlier \cite{LoewenO1990}.
The density field subsequently takes up the orientational inhomogeneity and both order parameters relax to their 
equilibrium profiles. The density develops a marked transient non-monotonic profile and relaxes much slower than the
orientational order. It takes quite a long time in units of the basic time scale of the dimensionless dynamical equations
\eqref{eq:PFCdI}-\eqref{eq:PhiQ} to end up in the final equilibrium state. 
Our finding shows that in principle our dynamical equations \eqref{eq:PFCdI} and \eqref{eq:PFCdII}, which reflect the diffusive dynamics of 
colloidal systems, can be applied to plenty of further growth phenomena in the future, which are, however, beyond the scope of the present paper.

\section{\label{sec:Zusammenfassung}Conclusions} 
In conclusion, we have explored the equilibrium structure of interfaces between various coexisting 
liquid crystalline phases using a PFC model for liquid crystals.
In two spatial dimensions, we have considered explicitly the isotropic-plastic crystalline and the 
smectic A-plastic crystalline interface, which are both anisotropic, \ie, they depend on the
relative orientation of the two coexisting phases.
To determine the equilibrium structures numerically, we calculated the relaxation of the dissipative PFC dynamics 
towards equilibrium (\ie, the minimization of the PFC functional) under the constant-mobility approximation 
using the finite element method. 

Basically, we have considered a two-order-parameter description of the interfaces containing the 
conserved (translational) density field and the non-conserved (orientational) nematic tensor. 
The phase diagram, the typical widths of the interfaces, the order-parameter profiles, and their dynamics were computed.
For the isotropic-plastic crystalline interface we find that in approaching the interface from the isotropic side, 
at first the nematic order builds up and then the density follows. 
The relative shift of the two profiles is about half the lattice constant of the plastic crystal. 
This finding is reminiscent to the fluid-crystal interface of systems of spherical particles 
\cite{Lutsko2010,HaertelOREHL2012}, which can be described by a two-order-parameter description involving the 
conserved mean density and a non-conserved crystallinity \cite{LoewenBW1989,LoewenBW1990}.
For the fluid-crystal interface, a similar shift has been found: if the interface is approached from the fluid side, 
first the (non-conserved) crystallinity increases and then the (conserved) mean density follows 
\cite{LoewenBW1989,LoewenB1990,OettelDBNS2012,Oettel2012}.
This has to do with the fact that a fluid is more responsive to an oscillatory density wave than to a 
global density change \cite{LoewenB1990}.

Our results can be verified either in particle-resolved computer simulations \cite{IvlevLMR2012} or in experiments. 
Particle-resolved computer simulations for rod-like systems have been performed both for structure
\cite{CleaverCAN1996,BolhuisF1997,AkinoSA2001,MarechalD2008} and dynamics \cite{Loewen1994b,KirchhoffLK1996} 
in various situations. 
So far experiments are concerned, most notably colloidal liquid crystals \cite{VroegeL1992,DogicF1997,LettingaDZMG2010} 
that are confined to two spatial dimensions are ideal realizations of our model.  
One important example is a suspension of the tobacco mosaic virus, which can be confined to monolayers \cite{WargackiPV2008} 
and which shows a variety of liquid crystalline phases \cite{GrafL1999b}, but there are more other examples
of liquid crystalline rod-like particle suspensions, which have been prepared in a controlled way 
(see, \eg, Refs.\ \cite{RoordavDPGvBK2004,TurkovicDF2005,CrassousDPMDHDS2012}).

Future work should extend the present study to three spatial dimensions \cite{WittkowskiLB2010,EmmerichEtAl2012}, 
which would require more numerical work but promises a richer equilibrium bulk phase diagram. 
Also the dynamics of a growing crystalline front, which has been studied for spherical particles already 
in detail \cite{SandomirskiALE2011,TegzeTG2011,RobbinsATK2012}, should be addressed for liquid crystals as well.
If a plastic crystalline phase grows into an isotropic phase, it would be interesting to follow the origin
of topological defects in the director field, which have to grow out of nothing. Moreover, crystal-fluid interfaces 
in external fields, like gravity, exhibit unusual effects already for isotropic particles \cite{BibenOL1994,AllahyarovL2011} 
and it would be challenging to explore this for liquid crystalline interfaces \cite{MarechalD2011}. Finally, our model
should be generalized to liquid crystals on manifolds \cite{NitschkeVW2012} to describe nematic \cite{DzubiellaSL2000} 
or smectic bubbles \cite{MayHTS2012}.

\appendix
\section{\label{A:Parameter}Notation}
Since the PFC model presented in Sec.\ \ref{sec:PFC} is equivalent to PFC models given in different notation in 
Refs.\ \cite{AchimWL2011,WittkowskiLB2011b,EmmerichEtAl2012}, we here clarify the relationship of our notation 
to the notation used in the literature. This especially simplifies the comparison of our phase diagram \ref{fig:pd} 
to the corresponding phase diagrams in Ref.\ \cite{AchimWL2011}.  

If we denote the eight parameters in Eqs.\ \eqref{eq:PFCs}, \eqref{eq:Jpsi}, and \eqref{eq:PhiQ} with a prime 
(\ie, $A_{1}'$, $A_{2}'$, $B_{3}'$, $D_{1}'$, $D_{2}'$, $\alpha'_{1}$, $\alpha'_{3}$, $\alpha'_{4}$) 
to avoid confusion with a similar notation in Refs.\ \cite{WittkowskiLB2011b,EmmerichEtAl2012}, 
the characteristic length $l'_{\mathrm{c}}$ and the characteristic energy $E'_{\mathrm{c}}$, 
which have been chosen to make the PFC model in the present article dimensionless, 
can be expressed by $l'_{\mathrm{c}}=\sqrt{-A_{3}/A_{2}}$ and $E'_{\mathrm{c}}=-(\pi\bar{\rho}/\beta)(A_{3}/A_{2})$ 
in terms of the reference particle number density $\bar{\rho}$, the inverse thermal energy $\beta$, 
and the parameters $A_{2}$ and $A_{3}$ in Ref.\ \cite{WittkowskiLB2011b}. 
In Ref.\ \cite{EmmerichEtAl2012}, the notation is analogously, but with $\rho_{\mathrm{ref}}$ instead of $\bar{\rho}$.  

The parameters in the free-energy functional \eqref{eq:PFCs} can be related to the parameters in 
Refs.\ \cite{WittkowskiLB2011b,EmmerichEtAl2012} by 
\begin{equation}
\!\;\,   A_{1}'=1-\frac{A_{1}}{2\pi\bar{\rho}}\;,\quad
A_{2}'=-\frac{A^{2}_{2}}{2\pi\bar{\rho}A_{3}}\;,\quad
B_{3}'=\frac{A_{2}B_{3}}{2\pi\bar{\rho}A_{3}}\;,   \!\!\!\!\!\!
\end{equation}
and
\begin{equation}
D_{1}'=\frac{1}{4}-\frac{D_{1}}{2\pi\bar{\rho}}\;,\quad
D_{2}'=\frac{A_{2}D_{2}}{2\pi\bar{\rho}A_{3}}\;.
\end{equation}
In case of the current \eqref{eq:Jpsi} and the quasi-current \eqref{eq:PhiQ}, a comparison with 
Refs.\ \cite{WittkowskiLB2011b,EmmerichEtAl2012} leads to the relations 
\begin{equation}
\alpha'_{1}=\frac{t_{\mathrm{c}}E_{\mathrm{c}}}{l^{2}_{\mathrm{c}}}\alpha_{1}\;,\quad\;
\alpha'_{3}=\frac{t_{\mathrm{c}}E_{\mathrm{c}}}{l^{2}_{\mathrm{c}}}\alpha_{3}\;,\quad\;
\alpha'_{4}=t_{\mathrm{c}}E_{\mathrm{c}}\:\!\alpha_{4}\;.
\end{equation}
In Ref.\ \cite{AchimWL2011}, a different notation is used. A comparison of the free-energy functional \eqref{eq:PFCs} 
with the corresponding free-energy functional in Ref.\ \cite{AchimWL2011} leads to 
\begin{equation}
A_{1}'=B_{l}\;,\quad
A_{2}'=4B_{x}\;,\quad
B_{3}'=-4F\;,\quad
\end{equation}
and
\begin{equation}
D_{1}'=2D\;,\quad
D_{2}'=8E\;.
\end{equation}
Furthermore, the length and energy scales of these two free-energy functionals are different. 
If $l'_{\mathrm{c}}$ and $E'_{\mathrm{c}}$ denote the characteristic length and energy in the present article and 
$l_{\mathrm{c}}$ and $E_{\mathrm{c}}$ denote the corresponding quantities in Ref.\ \cite{AchimWL2011}, 
they can be related to each other by 
\begin{equation}
l'_{\mathrm{c}}=\frac{1}{\sqrt{2}\:\!}\:\!l_{\mathrm{c}}\;,\quad
E'_{\mathrm{c}}=\frac{1}{2}E_{\mathrm{c}}\;.
\label{eq:vcglE}
\end{equation}

\acknowledgments{We thank Rainer Backofen for helpful discussions.  
R.W. gratefully acknowledges financial support from a Postdoctoral Research Fellowship (WI 4170/1-1) 
of the German Research Foundation (DFG).
This work was in addition supported by the DFG within SPP 1296.}

\bibliography{References}
\end{document}